\documentclass{aastex63}
\usepackage{epstopdf}
\begin{document}
\title{Satellite galaxies' drag on field stars in the Milky Way}

\correspondingauthor{Xilong Liang}
\email{xll@nao.cas.cn}

\author[0000-0001-9283-8334]{Xilong Liang}
\altaffiliation{School of Astronomy and Space Science, University of Chinese Academy of Sciences, Beijing 100049, China}
\altaffiliation{National Astronomical Observatories, Chinese Academy of Sciences,Beijing 100101, China}

\author{Jifeng Liu}
\altaffiliation{School of Astronomy and Space Science, University of Chinese Academy of Sciences, Beijing 100049, China}
\altaffiliation{National Astronomical Observatories, Chinese Academy of Sciences,Beijing 100101, China}
\altaffiliation{WHU-NAOC Joint Center for Astronomy, Wuhan University, Wuhan, China}

\author[0000-0003-2868-8276]{Jingkun Zhao}
\altaffiliation{National Astronomical Observatories, Chinese Academy of Sciences,Beijing 100101, China}
\altaffiliation{School of Astronomy and Space Science, University of Chinese Academy of Sciences, Beijing 100049, China}

\author{Kun Xu}
\altaffiliation{School of Astronomy and Space Science, University of Chinese Academy of Sciences, Beijing 100049, China}
\altaffiliation{National Astronomical Observatories, Chinese Academy of Sciences,Beijing 100101, China}

\begin{abstract}
With Gaia EDR3 data, velocity dispersion of Milky Way field stars around satellite galaxies have been investigated. We have fitted velocity dispersion against distance to satellite galaxy and found the gradient of velocity dispersion is related to the mass of satellite galaxy. With order-of-magnitude approximations, a linear correlation has been fitted between the mass of satellite galaxy and gradient of velocity dispersion caused by its gravitational drag. Though our result is an observational qualitative result, it shows better relation could be obtained with more observations in the future.

\end{abstract}

\section{Introduction}

The popular way to provide rough mass constraints for spheroidal galaxies is using the virial theorem \citep{1958BOTT....2q...3P}. With spherical Jeans equation, former researchers can estimate accurate masses for dispersion-supported stellar systems \citep{2010MNRAS.406.1220W}. Formalisms have been built between velocity dispersions of stars in dwarf galaxy and mass within its half-light radius, which can estimate accurate mass consistent with the result of full Jeans analysis \citep{2009ApJ...704.1274W}. As observational ability grows, more and more individual stars can be discerned within dwarf galaxies, but it is still easier to observe individual stars around the out skirt than those in the center of dwarf galaxies. Stars around boundary are more sensitive to traces of dynamical interaction between satellite galaxy and its host galaxy. Tidal tails are not target of this research and we focus on those satellites that show little evidence of Galactic cannibalism, namely no clear stellar streams striped from satellites by the tidal force of host galaxy.

Former researchers have investigated interaction of galaxies by application of dynamical friction formula \citep{1943ApJ....97..255C} or direct numerical simulation \citep{1983ApJ...264..364L,1983ApJ...274...53W}, but they mainly focused on orbital decay of satellite. \citet{1989MNRAS.239..549W} studied the response of a spherical stellar system to a periodic perturbation of satellite. \citet{1989ApJ...345..196C} found that sinking satellites can make the velocity dispersion horizontal to the disk larger than the vertical velocity dispersion, as is observed. \citet{1984MNRAS.208..687L} studied the influence of massive gas clouds on stellar velocity dispersions in galactic discs, which is close to our work. When a satellite galaxy passes through the halo of our Galaxy, impulse approximation is adopted to treat its gravitational effect on field stars. The impulse approximation \citep{1932PAAAS..67..169O} was introduced to calculate stellar perturbations on comets \citep{1950BAN....11...91O}, but the situation is similar for satellite galaxies.

Section 2 introduces the data and sample adopted to do this research. Section 3 introduces the model adopted to fit velocity distributions. The final section presents the result and discussion.

\section{Data analysis}

\subsection{data}
In the Gaia era \citep{2016A&A...595A...1G}, many stars around satellite galaxies have information available regarding spatial positions and proper motions.
For every satellite listed in Table 1, we selected all sources in Gaia EDR3 \citep{2021A&A...649A...1G} that are within $5^{\circ}$ from the center of each galaxy.
When those data are downloaded from the Gaia gaiaedr3.gaia\_source catalogue, some quality criteria have been adopted, which are respectively $\textrm{RUWE} \leq 1.4,$
$bp\_rp \geq 0.2 $ and $ parallax - parallax\_error < 0.03$, where RUWE is the renormalized unit weight error \citep{2021A&A...649A...3R}.
Parallaxes have been corrected using the prescription suggested in \citet{2021A&A...649A...2L} and used to select sample around each satellite galaxy. We did not use the parallax to calculate distance of each star, since most satellite galaxy are too far away. For each satellite galaxy, its surrounding stars have been selected by apply a parallax cut according to its known distance to Earth. The upper limits of parallax are all relatively tight, taken as about 10 to 30 kpc away from each different satellite galaxy. The lower limits of parallax are relatively loose that most of them are negative to include enough stars.
Then all stars within the parallax cut are assumed as the same distance as the satellite galaxies to Earth. Since we focus only on the distribution of tangent velocities, the error caused by the distance approximation will not be larger than an order of unity.
Velocities along right ascension and declination directions have been calculated by projecting all stars in the same distance as the satellite galaxy, which will no doubt induce huge uncertainties.
The fitting of velocity distribution can only be done in two dimensions without radial velocity and distance. Unfortunately, the sum of velocity dispersion along two directions is not necessarily two thirds of the total velocity dispersion in all three dimensions, because velocity dispersions may be different from each directions.
We think using sum of velocity dispersion in two dimensions representing that in three dimensions will only induces uncertainty with an order of unity.

To show variations of velocity distribution along the radius of each satellite galaxy, stars have been divided into spatial concentric annulus in the projected plane. The left panel of figure \ref{fig1} shows concentric annulus taken for Aquarius2 dwarf galaxy as an example.
These concentric annuluses have same area for the same satellite galaxy, but different areas for different galaxies, because the masses of different satellite galaxies and star densities of their field stars are different.
We assumed stars has approximately homogeneous distribution in space and the same area makes sure each annulus sample has similar numbers of stars for a same satellite galaxy.
Then the parallax approximation would has the same effect on all samples around one satellite galaxy.
Velocity distributions of stars in each annulus sample have been fitted separately.
The inner radius of smallest annulus of each satellite galaxy are taken larger than $0.5^{\circ}$, so that the velocity distribution is dominated by field stars belonging to the Milky Way. For each single star, whether it belongs to the satellite galaxy or field stars of the Milky Way is not important for us. What we want is the velocity distribution of field stars.
For each spatial annulus sample, we have divided it in velocity coordinate system into square bins with size equal 0.2 mas yr$^{-1}$ (about the median value of uncertainties of proper motion). Then took the star counts in each bin to represent the density distribution and took the mean value of uncertainties of proper motions of stars in each bin as uncertainty of that bin. We have ignored the uncertainty of distance which should be the dominating part of total uncertainty, but we think this way of treating uncertainty can still represent relative size of uncertainties between each bins. Then velocity distribution model described by equation \ref{eq1} has been adopted to fit the density distribution in velocity space. The EMCEE code  \citep{2013PASP..125..306F} has been used to obtain best fitted parameters and uncertainties of them.
We have used uniform priors on free parameters and the logarithm of probability function.
As for fitting steps, we first used the Python Scipy package optimize module to obtain least-squares estimation of parameters, then used them to initialize 100 random walkers. After threw away the initial 100 steps as "burn-in" steps, we run 1000 steps of MCMC. The parameters and their uncertainties are taken from 16th, 50th, and 84th percentiles of the samples in the marginalized distributions of MCMC results.

\subsection{model}

\begin{figure}
\plottwo{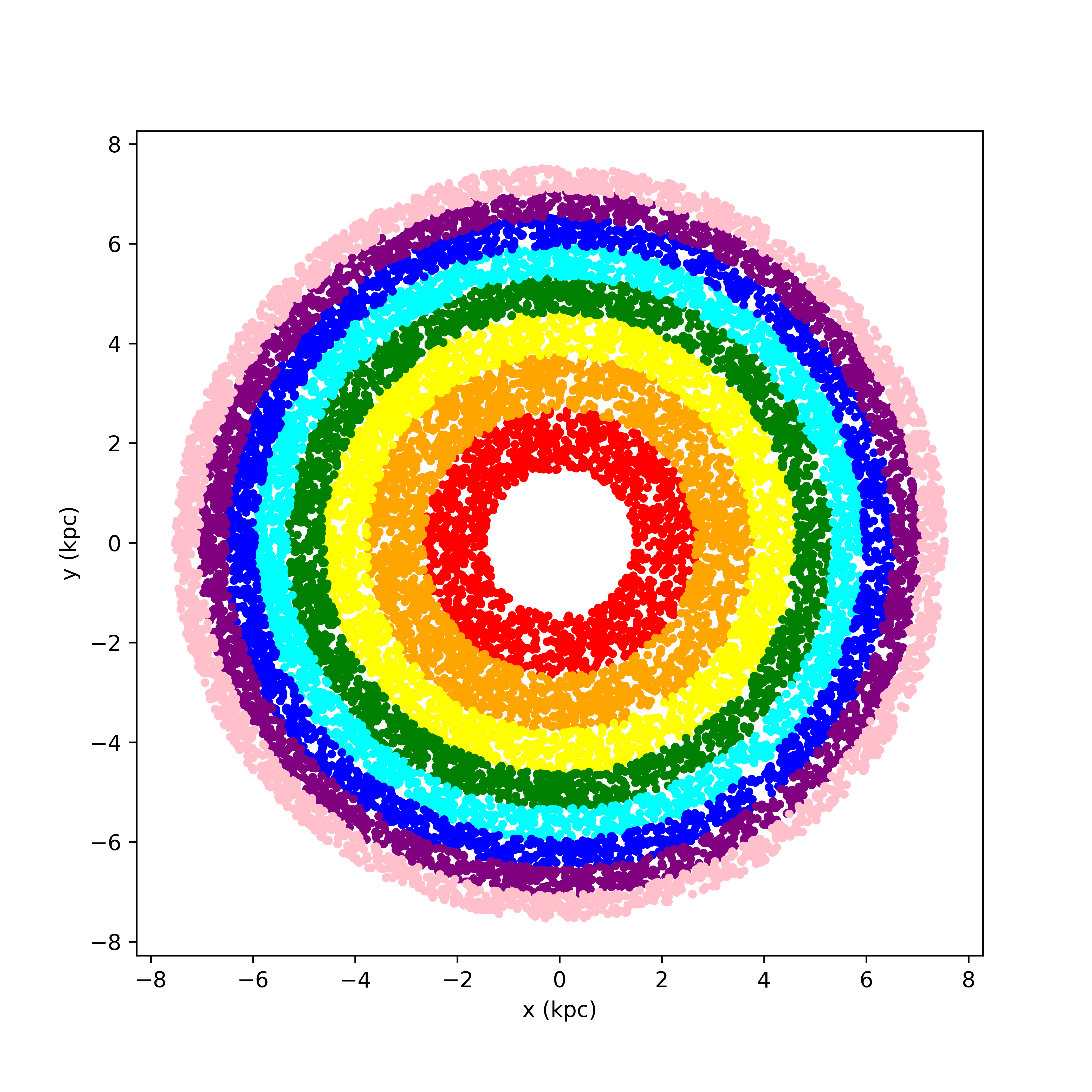}{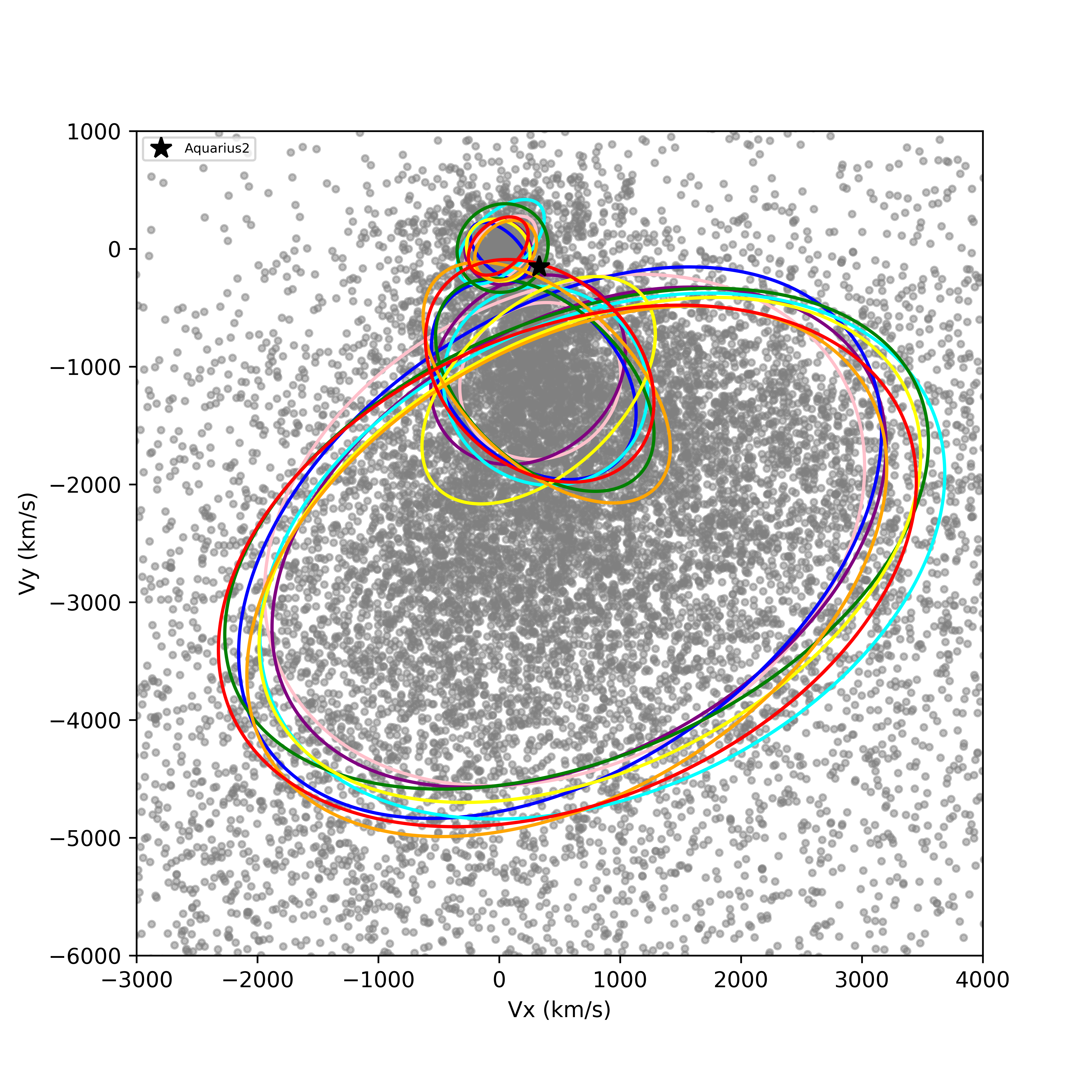}
\caption{The concentric annuluses taken for Aquarius2 dwarf galaxy and corresponding velocity eclipses of velocity distributions fitted for stars in each annulus. Each dot represent a star and colors are corresponding to each other in two panels. These grey dots in right panel are all stars in all annuluses, while the black star in the right panel is the known velocity of Aquarius2 dwarf galaxy from literature. \label{fig1}}
\end{figure}

The equilibrium state of an infinite, homogeneous stellar system can be described by Maxwellian function in phase space ($\overrightarrow{x}, \overrightarrow{v}$) and the velocity distribution is described by Schwarzschild distribution function.
We used three components to fit the velocity distributions of surrounding stars around satellite galaxies of the Milky Way, which are respectively for field stars from the Milky Way, stars from satellite galaxy itself and for possible observational artifact.
The right panel of figure \ref{fig1} shows fitted three components of velocity distributions of each annulus taken for Aquarius2 dwarf galaxy as an example.
The black star in the right panel shows the known velocity of Aquarius2 dwarf galaxy from literature \citep{2020AJ....160..124M}. Grey dots in the background show the velocity distribution of the whole sample. Three eclipses of each color represent three components fitted for stars in each annulus sample.
Though we do not know which star is actually from satellite galaxy, they are assumed having a distinct velocity distribution from field stars. With main part of stars from satellite galaxy already been fitted by one component, small number of stars from satellite galaxy mixed in the main part of field stars have negligible effect on the velocity distribution of field stars.

The Schwarzschild distribution function used for fitting velocity distribution is
\begin{equation}
f(v_1,v_2) = C exp(-\frac{((v_1-v_{1c})cos t+ (v_2-v_{2c})sin t)^2}{a^2}-\frac{-((v_1-v_{1c})sin t+ (v_2-v_{2c})cos t)^2}{b^2})
\end{equation}
in which $a$ and $b$ are two axes of the velocity ellipse, while $t$ is angular separation between axes of the velocity ellipse and axes of coordinate system. Coordinate $v = (v_1, v_2)$ is observed velocity of each star, while $(v_{1c}, v_{2c})$ is the center coordinate of the velocity ellipse. The same form of function has been applied to all three components and all parameters are free for fitting. We take the one fitted component with largest velocity dispersion as velocity distribution of field stars. The square of velocity dispersion is taken as $\sigma_v^2 = a^2 + b^2$.
We obtained velocity dispersions of the Milky Way field star component for every spatial concentric annulus. Then we fitted parameters for equation \ref{eq} with velocity dispersions against distance $r$ to the center of satellite galaxy. The EMCEE code has been used to obtain best fitted parameters and their uncertainties.
Though there are so many approximation treatment of uncertainties in velocity treatment, we think our result can still reveal some main tendencies.

The dynamical friction between a satellite galaxy and field stars of the Milky Way will systematically transfer energy from their relative orbital motion into random motions of their constituent stars. The process considered here is a satellite galaxy with mass $M$ passing through field stars with mass $m \ll M$. Though satellite galaxies are large complicated systems, we assume them as point masses for simplicity. Moreover, the spatial distribution of field stars has been roughly approximated as infinite and homogeneous, because the Galaxy is much larger than its satellite galaxies. While the satellite galaxy is much larger than field stars, the dominant effect of the encounters is to exert dynamical friction for the satellite galaxy.
For those stars with impact parameters smaller than radius of satellite galaxy, we assume they have been accreted by satellite galaxy or switch from one circular orbit to another and do not increase stars' random velocity. For those with impact parameters similar to radius of satellite galaxy, they experience strong encounters and we assume they have obtain much energy and distribute far away from the main part of stars in velocity coordinate system. Thus they do not affect the velocity distribution of the main part of field stars.

The impulse approximation is valid only if the encounter time is short compared to the crossing time. According to page 655, chapter 8.2 of \citet{2008gady.book.....B}, the crossing time (sometimes also called the dynamical time) can be estimated as $r/\sigma$, where $r$ is the Galactic radius and $\sigma$ is the velocity dispersion at the radius.
While the encounter time can be estimated as $\sim l/v$, where $l$ is the size of region passed through by satellite galaxy. Table \ref{apptab} lists our estimation of crossing time and encounter time of each satellite galaxy in our sample. $\sigma$ is calculated directly by square root of $\sigma_{v_0}^2$ listed in table \ref{apptab}. The size of region we are studying around satellite galaxy is around 10 kpc, which is used to calculate encounter time for the whole sample. To obtain relative velocity, fitted center of velocity distribution of Galaxy field star component is taken as tangential velocity of the Milky Way and its radial velocity is taken as zero. While for satellite galaxies, proper motions and radial velocities has been taken from \citet{2021ApJ...916....8L} and \citet{2019ARA&A..57..375S}. Coordinates module of Astropy package has been used to transform observable into velocities.
The impulse approximation is not a good approximation for some satellite galaxies that not have significantly shorter encounter time than crossing time and they have been removed from our sample.
With impulse approximation, the velocity perturbation in the encounter for each field star is
$$ \Delta v_{\perp} = \frac{GM}{bv}$$,
where G is the gravitational constant; $M$ is the mass of a satellite galaxy, which has been assumed as a point mass; $b$ is the distance of closest approach; while $v$ the velocity of a field star relative to the satellite galaxy. Supposing the direction of velocity of each field star does not change much during the impact, then the line of $b$ is approximately perpendicular to $\vec{v}$, thus

\begin{equation} \label{eq1}
b \thickapprox \frac{|\vec{r} \times \vec{v}|}{|\vec{v}|} = \frac{|(rcos \theta, rsin \theta) \times (v_1, v_2)|}{|\vec{v}|}.
\end{equation}

Where $\vec{r}=(rcos \theta, rsin \theta)$ is star's spatial coordinate with origin at the satellite galaxy, while $\vec{v} = (v_1, v_2)$ is star's velocity coordinate. We made a simplified approximation that the encounter only adds energy along the direction perpendicular to the original velocity direction. Since stars has been assumed homogeneous distribution in space, the mean of velocity perturbation is zero, but the velocity dispersion becomes larger.
The variance of perturbed velocity component $v_{\perp}$ can be calculated by integrating over entire phase space:

\begin{eqnarray}
\overline{(\Delta v_{\perp})^2} &=& \int(\frac{GM}{bv})^2f(v) dv dr rd\theta  \\
~~~~~~~~~~~~~~~~~~~~~~~ &=& \int\frac{G^2 M^2}{(v_2 r cos \theta - v_1 r sin \theta)^2} f(v_1, v_2) dr dv_1 dv_2 rd\theta \\
~~~~~~~~~~~~~~~~~~~~~~~ &=& \int\frac{G^2 M^2}{r} dr\frac{f(v_1, v_2) dv_1 dv_2 d\theta}{(v_1 sin \theta - v_2 cos \theta)^2} \label{eq5}
\end{eqnarray}

The $M$ is constant, while $$ \int G^2\frac{f(v_1,v_2)dv_1 dv_2 d\theta}{(v_1 sin \theta - v_2 cos \theta)^2}$$ integrated over the whole velocity space is only a function of velocity dispersion and has little dependency on $r$. If denoting the integral by $C$, the equation \ref{eq5} can be expressed as

\begin{equation}
\overline{(\Delta v_{\perp})^2} = M^2C\int\frac{dr}{r}
\end{equation}

\begin{figure}
\plotone{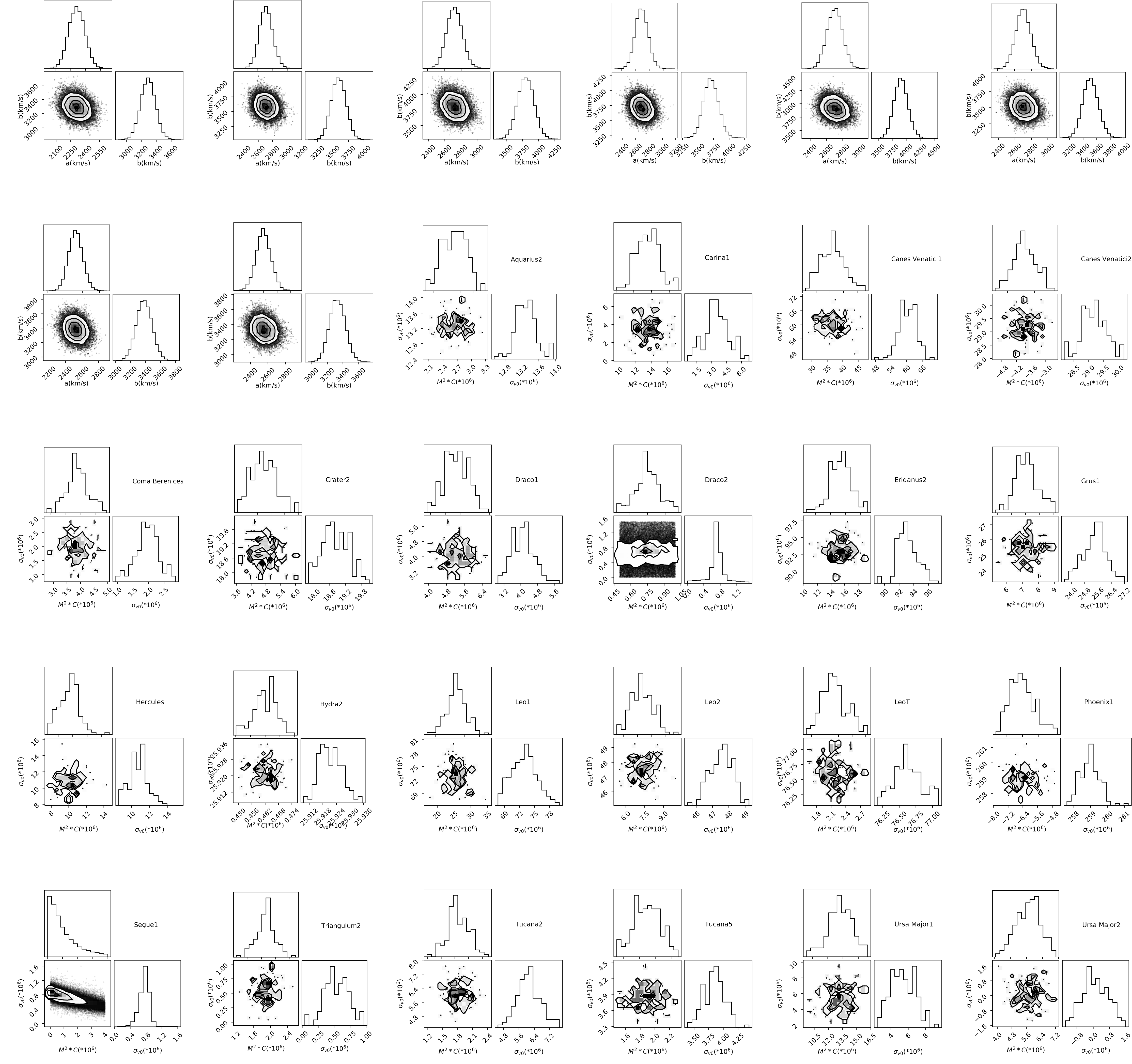}
\caption{The corner plots of fitted two axes of velocity ellipses of annulus samples of AquaII and $M^2C$ versus $\sigma_{v_0}^2$ for each satellite galaxy. \label{corner}}
\end{figure}

For an annulus with small width, the change of velocity square is approximately:

\begin{equation}\label{eq}
\overline{(\Delta v_{\perp})^2} = (\ln \frac{r+\Delta r}{r}) M^2C = \sigma_v^2 - \sigma_{v0}^2
\end{equation}

where $C$ is a constant to $r$ under condition ignoring possible correlation between velocity distribution and spatial distribution of stars in an annulus.
The $\sigma_v^2$ is the present total velocity dispersion, while $\sigma_{v_0}^2$ is the total velocity dispersion before velocity perturbation. Fitting $\sigma_v^2$ against $(\ln \frac{r+\Delta r}{r})$, we have obtained $M^2C$ and $\sigma_{v_0}^2$ at the same time.
The first eight panels of figure \ref{corner} show corner distributions in parameter space of two axes of the velocity ellipse for each annulus sample of AquaII. While the rest panels show corner distributions of $M^2C$ versus $\sigma_{v_0}^2$ of each satellite galaxy in our sample. Though only a few points have been used to fit equation \ref{eq} for each satellite galaxy, we think it is enough to reveal the main relation.

\section{Result and Discussion}

\begin{figure}[hbtp]
\plotone{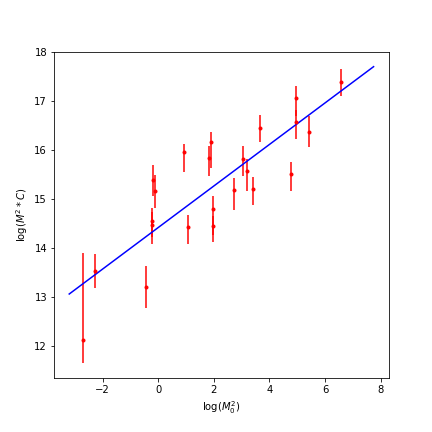}
\caption{Fitted $M^2C$ versus known mass $M_0^2$. \label{fig}}
\end{figure}

Figure \ref{fig} shows fitted $M^2C$ versus known mass $M_0^2$ in logarithmic coordinate which is for aesthetics since mass square spans several orders of magnitude. $M^2C$ is fitted coefficients from equation \ref{eq}, while $M_0^2$ is known masses of satellite galaxies from literatures listed in table \ref{apptab}. The blue straight line is only a line fitted between $M^2C$ and $M_0^2$ and does not indicate $C$ has to be a constant. The error bars symbolically show three times of uncertainties obtained by former steps, since uncertainty brought by distance should be much larger than uncertainty brought by proper motion.
The spearman correlation coefficient between $M^2C$ and $M_0^2$ is 0.81, which means their correlation is close to linear relation. The spearman correlation coefficient between $\sigma_{v0}^2$ and distance from Earth is 0.96, which means their correlation is also very close to linear relation. Our result at least qualitatively shows that gradient of velocity dispersion of field stars around a satellite galaxy is strongly related to the total mass of the satellite galaxy. Without full three dimensional velocity information, it is meaningless to use more complicated functions to describe the correlation than linear relation.

Since gravity is always an attractive force, a equilibrium gravitational system must be inhomogeneous. Our relation is obtained under the assumption of an infinite, homogeneous stellar system of field stars which is only an approximation of the real observed spatial distribution, but can illuminate much of the response behavior of real stellar system. We think the total uncertainty should not be more than an order of magnitude larger than parameter itself. The gravitational mass of satellite galaxy is no doubt related to the enlargement of velocity dispersion of field star caused by its gravitational drag.

Our result is only an order-of-magnitude argument, but in the future, large surveys like CSST and LSST will cover more stars in local universe and they can provide radial velocity and even chemical abundances information. With more proper motions and parallaxes information from Gaia future data release, it is possible that much more stars will have full 6 dimensional information. Even then distinguishing single stars in the dense center of dwarf galaxies is still very hard, more precise analysis about field stars around satellite galaxy is promising.

\acknowledgments

We thank the anonymous referee for his/her constructive suggestions. This work is supported by Strategic Priority Program of the Chinese Academy of Sciences under grant number XDB41000000, the Fundamental Research Funds for the Central Universities and Project funded by China Postdoctoral Science Foundation No. 2021M703168. This study is also supported by the National Natural Science Foundation of China under grant No. 11973048, 11927804, 11890694 and National Key R\&D Program of China No. 2019YFA0405502. We acknowledge the support from the 2m Chinese Space Station Telescope project: CMS-CSST-2021-A10, CMS-CSST-2021-B03, CMS-CSST-2021-B05.

The authors also acknowledge all the open-source software involved in this study, specially TOPCAT, Python and R soft. This work has made use of data from the European Space Agency (ESA) mission Gaia (https://www.cosmos.esa.int/gaia), processed by the
Gaia Data Processing and Analysis Consortium (DPAC, https://www.cosmos.esa.int/web/gaia/dpac/consortium). Funding for the DPAC has been provided by national institutions, in particular, the institutions participating in the Gaia Multilateral Agreement.

\appendix
Table \ref{apptab} lists fitted parameters in equation \ref{eq}, distances, known masses of satellite galaxies and referenced literatures. Column $ct$ and $ et $ are  respectively crossing time and encounter time in unit of million year.

\begin{table}
\begin{center}
\caption{Parameters.\label{apptab}}
\begin{tabular}{llllllll}
\hline
\hline
Name              & $M^2C$                  & $\sigma_{v0}^2$    &$dis$ & $M_0$ &$ct  $ &  $ et $&       reference\\
\hline
unit     & $10^{4}$ & $10^{4}$ km$^2$s$^{-2}$& kpc & $10^{6}M_{sun}$ & Myr &Myr &~ \\
\hline

Aquarius2      &$ 267   ^{+26  }_{ -37}$&  $ 1332 ^{+28  }_{ -23}$ & 105  & 2.7  &  28.8 &   4.0 &\citet{2016MNRAS.463..712T} \\
Carina1        &$ 1390  ^{+134 }_{-117}$&  $  363 ^{+123 }_{-126}$ & 107  & 6.3  &  55.8 &   5.1 &\citet{2012AJ....144....4M} \\
Canes Venatici1&$ 3561  ^{+337 }_{-305}$&  $ 5995 ^{+412 }_{-321}$ & 218  & 27.0 &  28.1 &   2.7 &\citet{2007ApJ...670..313S} \\
Canes Venatici2&$ 480   ^{+58  }_{-44 }$&  $ 2811 ^{+49  }_{ -41}$ & 161  & 0.91 &  30.2 &   4.0 &\citet{2012AJ....144....4M} \\
Coma Berenices &$ 378   ^{+48  }_{-37 }$&  $  197 ^{+39  }_{ -33}$ & 45   & 0.94 &  32.1 &   8.6 &\citet{2012AJ....144....4M} \\
Crater2        &$ 397   ^{+38  }_{-37 }$&  $ 1871 ^{+43  }_{ -39}$ & 116  & 5.5  &  26.7 &   4.5 &\citet{2021ApJ...921...32J} \\
  Draco1       &$ 539   ^{+53  }_{-51 }$&  $  409 ^{+54  }_{ -47}$ & 76   & 11.0 &  37.5 &   8.8 &\citet{2012AJ....144....4M} \\
  Draco2       &$ 75    ^{+10  }_{-7  }$&  $   71 ^{+8   }_{  -7}$ & 24   & 0.32 &  28.7 &  12.4 &\citet{2016MNRAS.458L..59M} \\
  Eridanus2    &$ 1551  ^{+146 }_{-148}$&  $ 9261 ^{+146 }_{-162}$ & 382  & 12.0 &  39.6 &   2.6 &\citet{2017ApJ...838....8L} \\
  Grus1        &$ 755   ^{+65  }_{-80 }$&  $ 2546 ^{+75  }_{ -71}$ & 116  & 2.5  &  23.0 &   4.9 &\citet{2016ApJ...819...53W} \\
  Hercules     &$ 103   ^{+82  }_{-140}$&  $ 1081 ^{+99  }_{-121}$ & 126  & 2.6  &  38.4 &   3.3 &\citet{2012AJ....144....4M} \\
  Hydra2       &$ 54    ^{+9   }_{-6  }$&  $ 2575 ^{+8   }_{  -5}$ & 148  & 0.8  &  29.1 &   3.1 &\citet{2015ApJ...810...56K} \\
  Leo1         &$ 2550  ^{+229 }_{-256}$&  $ 7270 ^{+202 }_{-194}$ & 258  & 12.0 &  30.2 &   1.4 &\citet{2012AJ....144....4M} \\
  Leo2         &$ 734   ^{+76  }_{-71 }$&  $ 4770 ^{+67  }_{ -79}$ & 236  & 4.6  &  34.0 &   1.6 &\citet{2012AJ....144....4M} \\
  LeoT         &$ 393   ^{+34  }_{-43 }$&  $ 7647 ^{+37  }_{ -40}$ & 422  & 3.9  &  48.2 &   1.5 &\citet{2012AJ....144....4M} \\
  Phoenix1     &$ 205   ^{+22  }_{-18 }$&  $25783 ^{+22  }_{ -21}$ & 409  & 0.89 &  25.4 &   0.9 &\citet{2012AJ....144....4M} \\
  Pisces2      &$ 841   ^{+48  }_{-94 }$&  $ 3889 ^{+95  }_{ -90}$ & 181  & 1.6  &  29.0 &   3.2 &\citet{2015ApJ...810...56K} \\
  Segue1       &$  18   ^{+30  }_{-2  }$&  $   84 ^{+2   }_{  -5}$ & 28   & 0.26 &  30.3 &   5.9 &\citet{2012AJ....144....4M} \\
  Triangulum2  &$ 192   ^{+19  }_{-20 }$&  $   53 ^{+18  }_{ -21}$ & 36   & 0.89 &  49.4 &  20.6 &\citet{2015ApJ...814L...7K} \\
  Tucana2      &$ 183   ^{+15  }_{-17 }$&  $  619 ^{+52  }_{-54 }$ & 54   & 2.7  &  21.8 &  10.2 &\citet{2016ApJ...819...53W} \\
  Tucana5      &$ 184   ^{+16  }_{-18 }$&  $  385 ^{+16  }_{ -18}$ & 52   & 1.7  &  26.4 &   9.7 &\citet{2020ApJ...892..137S} \\
  Ursa Major1  &$ 1276  ^{+166 }_{-110}$&  $  519 ^{+134 }_{-126}$ &  97  & 15   &  42.5 &   6.3 &\citet{2007ApJ...670..313S} \\
  Ursa Major2  &$ 574   ^{+53  }_{-64 }$&  $   10 ^{+67  }_{ -47}$ &  32  & 4.9  &  97.8 &  28.5 &\citet{2007ApJ...670..313S} \\

\hline
\end{tabular}
\end{center}
\end{table}

\end{document}